\documentclass[3p,times,11pt]{elsarticle}
\usepackage{pifont}
\usepackage{amsmath, amsthm, amssymb, amsfonts}
\usepackage{graphicx}
\usepackage{geometry}
\usepackage{txfonts}
\usepackage[colorlinks]{hyperref}
\usepackage{natbib}
\usepackage[T1]{fontenc}
\usepackage[all]{xy}
\usepackage{times}
\usepackage{yfonts}
\usepackage[utopia]{mathdesign}
\usepackage{caption}
\usepackage{algorithm}
\usepackage{color}
\usepackage[noend]{algorithmic} 
\algsetup{indent=3mm} 

\newtheorem{Theorem}{Theorem}
\newtheorem{Example}{Example}[section]
\newtheorem{Remark}{Remark}

\newtheorem{Definition and Notation}{Definition and Notation}
\newtheorem{Definition}{Definition}[section]
\newtheorem{Lemma}{Lemma}

\newtheorem{Proposition}{Proposition}
\newtheorem{Corollary}{Corollary}
\newenvironment{Proof}[1][Proof]{\noindent {\bf Proof.\;}}{\qed}

\journal{--}

\usepackage{amssymb}

\usepackage[figuresright]{rotating}

\begin{document}

\begin{frontmatter}

\title{Do non-free LCD codes over finite commutative Frobenius rings exist?}


\author[sb]{Sanjit Bhowmick}\ead{sanjitbhowmick392@gmail.com}
\author[af]{A. Fotue-Tabue}\ead{alexfotue@gmail.com}
\author[em]{E. {Mart\'inez-Moro}}\ead{ {edgar.martinez@uva.es}}
\author[rb]{Ramakrishna Bandi\corref{cor1}}\ead{ramakrishna@iiitnr.edu.in}\cortext[cor1]{Corresponding author.}
\author[sb]{Satya Bagchi}\ead{satya.bagchi@maths.nitdgp.ac.in}
\address[sb]{Department of Mathematics, National Institute of Technology Durgapur, Durgapur, India}
\address[af]{Department of Mathematics, Faculty of Science, University of Yaoundé I, Cameroon}
\address[rb]{Department of Mathematics, Dr. SPM International Institute of Information Technology, Naya Raipur, India}
\address[em]{Institute of Mathematics, University of Valladolid, Spain}

\begin{abstract} In this paper, we clarify some aspects on LCD codes in the literature. We first prove that a non-free LCD code does not exist over finite commutative Frobenius local rings. We  then obtain a necessary and sufficient condition for the existence of  LCD code over finite commutative Frobenius rings. We later show that a free constacyclic code over finite chain ring is LCD if and only if it is reversible, and also provide a necessary and sufficient condition for a constacyclic code to be reversible over finite chain rings. We illustrate the  minimum Lee-distance of  LCD codes over some finite commutative chain rings and demonstrate the results with examples. We also got some new optimal $\mathbb{Z}_4$ codes of different lengths {which are} cyclic LCD codes over $\mathbb{Z}_4$. 
\end{abstract}


\end{frontmatter}



\section{Introduction}

Codes over finite rings is  quite a popular topic of interest.  A
linear code $C$  a su is called {\it{Linear Complementary Dual (LCD) code}} if
$C$ meets its dual $C^\perp$ trivially.  LCD codes  were first
investigated by Massey{, he showed there  a characterization of} LCD codes
and non-LCD codes over finite fields and demonstrated that
asymptotically good LCD codes exist \cite{Mas92}. LCD codes have
been widely applied in data storage, communications systems,
consumer electronics, and cryptography. Carlet and Guilley shown
an application of LCD codes against side-channel attacks and fault
injection attacks, and presented several constructions of LCD
codes \cite{CG16}.  Cyclic LCD codes over finite fields  are
also referred as reversible codes, Yang and
Massey gave a necessary and sufficient condition for a cyclic code
to have a complementary dual \cite{YM94} and proved that reversible
 cyclic codes over finite fields are LCD codes. In \cite{Mas64}, Massey showed
that some cyclic LCD codes over finite fields are BCH codes, and
also constructed reversible convolutional codes which are in fact LCD codes.  
 Tzeng and Hartmann \cite{TH70}  proved that the minimum distance of a class of LCD
codes is greater than that given by the BCH bound. Using the hull
dimension spectra of linear codes, Sendrier showed that LCD codes
meet the asymptotic Gilbert-Varshamov bound \cite{Sen04}.
Dougherty et. al. developed a linear programming bound on the
largest size of an LCD code of given length and minimum distance
\cite{DKOSS}. Guneri et. al. studied quasi-cyclic complementary
dual codes using their concatenated structure in \cite{GOS16} and
\cite{GOOSSS17}. Ding et al. constructed several families of
cyclic LCD codes over finite fields and analyzed their parameters
\cite{DLL17}. In \cite{LDL}, Li et al. studied a class of LCD BCH
codes. Jin showed that some Reed-Solomon codes are equivalent
to LCD codes \cite{Jin2016}. In \cite{CMT18}, the authors
proved that any MDS code is equivalent to an LCD code and
constructed LCD Maximum distance Separable codes.  Jitman
et. al. studied Complementary dual subfield linear codes over
finite fields \cite{BJ16}. 

Recently, in \cite{LL15}, Liu and Liu studied LCD
codes over finite chain rings and provided a necessary and
sufficient condition for a free linear code to be a LCD code over
finite chain ring. They also gave a sufficient condition for a
linear code (not necessarily free) over a finite chain ring to be
LCD code, which says "A linear code $C$ over a finite chain ring
with generator matrix $G$ is LCD code if $\mathrm{G}\mathrm{G}^T$
is invertible, where $G^T$ is the transpose of $G$" \cite[Theorem
3.5]{LL15}. They provided an example \cite[Example 2]{LL15} to
state that the converse of \cite[Theorem 3.5]{LL15}  is not in
general true. However there is a mistake in their example. In this
paper, we prove that the converse of \cite[Theorem 3.5]{LL15} is
indeed true. This  lead to the main result (see Theorem\,\ref{thm1}) of this 
paper,  it proves that there are no non-free LCD codes over finite commutative local
Frobenius rings  by showing  that any LCD code over a finite
commutative Frobenius ring is the Chinese product of LCD codes
over finite commutative local Frobenius rings (see
Theorem\,\ref{thm}). The other contributions are the characterizations
of projection of LCD codes (see Theorem\,\ref{thm:proj1} ) and lift
LCD codes (see Theorem \ref{thm:proj2}) over a finite commutative
local Frobenius ring. We also show
that a free constacyclic code $C$ over finite chain ring is LCD code if
and only if $C$ is reversible.   We also prove a necessary and sufficient
condition for a constacyclic code $C$ of length $n$ over finite chain
rings to be reversible when $n$ is relatively prime to the
characteristic of the finite chain ring.

 The paper is organized as follows: In Section \ref{sec:2}, we provide
some basic tools which are required to understand the results of
further sections. In Section \ref{sec:3}, we discuss LCD codes
over finite commutative Frobenius rings. Finally, Section
\ref{sec:4} studies the complementary dual constacyclic codes
over  finite chain rings in more general setting by a uniform
method.

\section{Some notations and basic results of codes over finite commutative  Frobenius
rings}\label{sec:2}

Throughout this section $R$ is a commutative finite ring with
multiplicative unity $1$ distinct to $0.$   A commutative finite
ring $R$ is Frobenius if  $R$ as $R$-module is injective.
Alternatively, we can say a finite ring is Frobenius if
$R/\texttt{J}(R)$ is isomorphic to $\texttt{soc}(R)$ (as
$R$-modules), where $\texttt{J}(R)$ is the Jacobson radical 
 and $\texttt{soc}(R)$ is the socle of the ring $R$. Recall
that the Jacobson radical is the intersection of all maximal
ideals in the ring and the socle of the ring is the sum of the
minimal $R$-submodules. A ring is a local ring if it has  unique
maximal ideal. A principal ideal ring is a ring such that each of its 
ideal is generated by a single element.

Let $R$ be a finite ring with maximal ideals
$\textgoth{m}_1,\cdots,\textgoth{m}_u$ and $s_1,\cdots,s_u$ their
indices of stability, respectively. Clearly
$R/\textgoth{m}_j^{s_j}$ is a finite local ring  with maximal
ideal $\textgoth{m}_j/\textgoth{m}_j^{s_j}.$  Then we have the
ring epimorphisms
\begin{align}
\begin{array}{cccc}
 \Phi_j: & R & \rightarrow & R_j \\
    & a & \mapsto & a+\textgoth{m}_j
\end{array}
\end{align} and
$\texttt{Ker}(\Phi_j)=\textgoth{m}_j^{s_j}$ ($1\leq j\leq u$). The
ring epimorphisms $\Phi_j$ ($1\leq j\leq u$) induce the following
ring homomorphisms
\begin{align}\label{Phi}
\begin{array}{cccc}
 \Phi: & R & \rightarrow & R_1\times\cdots\times R_u \\
    & a & \mapsto & \left(\Phi_1(a),\cdots,\Phi_u(a)\right).
\end{array}
\end{align}
 Since the maximal ideals
$\textgoth{m}_1,\cdots,\textgoth{m}_u$ of $R$ are coprime and
$\bigcap_{j=1}^u \textgoth{m}_j^{s_j} = \{0_R\},$ the ring
homomorphism (\ref{Phi}) is a ring isomorphism, by the Chinese
remainder theorem, see \cite[p.224]{McD74}. We denote the inverse
of this map by $\texttt{CRT}$ and we say that $R$ is the
\emph{Chinese product of rings} $\{R_j\}_{j=1}^u.$ 

\begin{Theorem} \cite[p.224]{McD74} Let $R$ be a Frobenius ring, then $$R = \texttt{CRT}(R_1, R_2, \cdots, R_u),$$ where $R_j$ is a local Frobenius ring.
\end{Theorem}
That is $R_j := R/\textgoth{m}_j^{s_j}$ is a local Frobenius ring
for each $j.$ The following is an example of a ring that is
local Frobenius ring but not a chain ring. We shall use this ring
to exhibit several of the results of the paper.

\begin{Example}\cite{DYK11} Let $R_m = \mathbb{Z}_2[u_1,u_2,\cdots,u_m]/\langle\,u_1^2,u_2^2,\cdots,u_m^2\,\rangle,$
where $\langle\,u_1^2,u_2^2,\cdots,u_m^2\,\rangle$ denotes the
ideal generated by $u_1^2,u_2^2,\cdots,u_m^2,$ and $m$ is a
positive integer. Then $$\texttt{J}(R_m)
=\langle\,u_1,u_2,\cdots,u_m\,\rangle\hbox{ and }\texttt{Soc}(R_m) =\left\langle\,\prod\limits_{i=1}^k
u_i\,\right\rangle.$$ Thus $R_m/\texttt{J}(R_m) \cong
\texttt{Soc}(R_m)$ (as $R_m$-modules), so $R_m$ is a finite
commutative local Frobenius ring. However $R_m$ is non-chain if
 $m>1.$
\end{Example}

We shall use the previous decomposition of rings to understand codes
defined over finite commutative  Frobenius rings. The zero element
in $R^n$ will be denoted as $\textbf{0}.$ A linear code $C$ of length $n$
over a finite ring $R$ is an $R$-submodule of $R^n.$ Let $C_j$ be
a code of length $n$ over $R_j,$ and  extend the map $ \Phi$
coordinatewise to $R^n$ as
\begin{align}
\begin{array}{cccc}
 \Phi: & R^n & \rightarrow & (R_1)^n\times\cdots\times(R_u)^n \\
    & \textbf{a} & \mapsto &
    \left(\Phi_1(\textbf{a}),\cdots,\Phi_u(\textbf{a})\right),
\end{array}
\end{align}
where
\begin{align}\label{Phi}
\begin{array}{cccc}
 \Phi_j: & R^n & \rightarrow &  (R_j)^n \\  & (a_1,a_2,\cdots,a_n) & \mapsto &
 \left(\Phi_j(a_1),\Phi_j(a_2),\cdots,\Phi_j(a_n)\right).
\end{array}
\end{align}
Then  $C = \texttt{CRT}(C_1,C_2, \cdots , C_u) =  \Phi^{-1}(C_1 \times
C_2 \times \cdots \times C_u),$ where $\Phi_j(C)=C_j$ for $1\leq
j\leq u.$ We say that $C$ is the Chinese product of codes
$C_1,C_2, \cdots ,C_u.$  This allows us to reduce the study of
codes over finite commutative Frobenius rings to that of codes
over finite commutative local Frobenius rings. The rank of a
linear code $C$ over $R$ of length $n,$ is defined by
$$\texttt{rank}_R(C):= \texttt{min}\left\{i\in\mathbb{N} \;:\;
\text{there exists a monomorphism $C \hookrightarrow R^i$ as
$R$-modules}\right\}.$$  We say that a linear code $C$ over $R$ is
free if $C$ is isomorphic (as a module) to $R^t$ for some $t.$ It is
immediate that if $C$ is free then $\texttt{rank}_R(C) = t,$ where
$C \cong R^t.$ A linear $[n,k]$-code over $R$ is an $R$-submodule
of $R^n$ of rank $k.$  Note that a standard generator matrix for
any free linear $[n,k]$-code $C$ over $R$ is of the form
$[\mathrm{I}_k\,|\, \mathrm{M}]\mathrm{U},$ where $M$ is a matrix
over $R,$ $\mathrm{U}$ is a permutation matrix and
$k=\texttt{rank}_R(C).$

\begin{Lemma}\cite[Theorem 2.4]{DL09}\label{lem3} Let  $C_j$ be a linear code  over $R_j$ for $i = 1, 2, \ldots , u,$ and 
$C = \texttt{CRT}(C_1,C_2, \cdots, C_u).$ Then
\begin{enumerate}
    \item $|C| = \prod_{i=1}^u|C_j|;$
    \item $\texttt{rank}_R(C)=\texttt{max}\left\{\texttt{rank}_{R_j}(C_j)\;:\;1\leq j\leq u\right\};$
    \item  $C$ is a free code if and only if each $C_j$ is a free code with the
    same rank  $\texttt{rank}_R(C).$
\end{enumerate}
\end{Lemma}

We attach the standard inner-product to  $R^n$, that is
 \begin{align}[\textbf{v}\,,\,\textbf{w}] =
\sum_{j=1}^n v_jw_j,\end{align} where
$\textbf{v}:=(v_1,v_2,\cdots,v_n)$ and
$\textbf{w}:=(w_1,w_2,\cdots,w_n)$ are elements in $R^n.$ 
 For a code $C,$ its dual code is defined as
follows:
\begin{align}C^\perp = \{\textbf{u}\in R^n \;:\; [\textbf{u},
\textbf{c}] = 0_R, \text{ for all } \;\textbf{c}\; \text{ in }
C\}.\end{align} It is well known that for codes over Frobenius
rings, $|C||C^\perp| = |R|^n,$ (see \cite{Woo99} for a proof).

\begin{Lemma}\label{lem4}\cite[Theorem 2.7]{DL09}
If $C = \texttt{CRT}(C_1,C_2, \cdots , C_u )$ is a code over $R,$
then $C^\perp = \texttt{CRT}(C_1^\perp ,C_2^\perp , \cdots,
C_u^\perp).$
\end{Lemma}

For the rest of the paper  $R$ will denote  the Chinese product of finite commutative local  Frobenius rings $R_1,R_2,\ldots, R_u$ unless otherwise is specified. Let
$\mathrm{M}_{k\times n}(R_j)$ be the set of all $k\times
n$-matrices over $R_j.$ For $\mathrm{A}\in\mathrm{M}_{k\times
n}(R_j),$ the transpose of the matrix $\mathrm{A}$ is denoted by
$\mathrm{A}^T.$ We also let $\textbf{0}$ denote the zero matrix,
where the size will either be obvious from the context or
specified whenever necessary. Similarly, we denote the $k\times k$
identity matrix by $I_k.$ The elements $\textbf{c}_1, \textbf{c}_2, \ldots,
\textbf{c}_k\in R^n$ are said to be \emph{linearly independent} over
$R_j$ if for all $(x_1, x_2,\cdots, x_k)$ in the set $(R_j)^k$ such
that $x_1\textbf{c}_1+ x_2\textbf{c}_2+\cdots+ x_k\textbf{c}_k= \textbf{0}$
implies that $x_1= x_2= \cdots = x_k=0.$ If the rows of a $k\times
n$-matrix $\mathrm{A}$ over $R_j$ are linearly independent, then
we say that $\mathrm{A}$ is a \emph{full-row-rank} matrix. If
there is an $k\times n$-matrix $\mathrm{B}$ over $R_j$ such that
$\mathrm{A}\mathrm{B} = \mathrm{I}_k,$ then we say that
$\mathrm{A}$ is \emph{right-invertible} and $\mathrm{B}$ is a
right inverse of $\mathrm{A}.$ When $k = n,$ we say that
$\mathrm{A}$ is non-singular, if the determinant
$\texttt{det}(\mathrm{A})$ is a unit of $R_j.$ Otherwise,
$\mathrm{A}$ is said to \emph{singular}. Note that a matrix
$\mathrm{A}$ is invertible over $R_j,$ if and only if $\mathrm{A}$
is nonsingular over $R_j.$ The following two results about
full-row-rank matrices over finite commutative Frobenius rings
appear in \cite{FLL14}.

\begin{Lemma}\label{lem:full} Let $R_j$ be a finite commutative Frobenius rings.
A $k\times n$-matrix $\mathrm{A}$ is full-row-rank, if and only if
$\mathrm{A}$ is right-invertible.
\end{Lemma}

\begin{Lemma}  Let $\mathrm{A}$ be an $k\times k$-matrix over a finite commutative Frobenius ring
$R.$ The following statements are equivalent:
\begin{enumerate}
    \item  $\mathrm{A}$ is invertible.
    \item  $\mathrm{A}$ is non-singular.
    \item $\mathrm{A}$ is full-row-rank.
\end{enumerate}
\end{Lemma}

The next corollary follows from a typical linear algebra argument.

\begin{Corollary} \label{cor:ax}  The $k\times n$-matrix $\mathrm{A}$
over $R_j$ is singular, if and only if there is a nonzero vector
$X:=(x_1, \cdots, x_k)^T$ in $R^k$ such that $\mathrm{A}X =
\textbf{0}.$
\end{Corollary}

\section{Characterization of LCD codes over finite commutative Frobenius rings}\label{sec:3}

 In \cite[Theorem 3.5]{LL15}, it is proved that any linear code $C_j$ over $R_j$
with a generator matrix $\mathrm{G}_j$ is LCD if,
$\mathrm{G}_j\mathrm{G}_j^T$ is invertible, and other
hand it is also stated that the converse of \cite[Theorem 3.5]{LL15} is
not true in general with an example \cite[Example 2]{LL15}. However there is a mistake in that example (as $(2,0,0,2,0)$ is in $C\cap C^\perp$).  From \cite[Corollary 3.6.]{LL15}, if $C_j$ is free then the
converse of \cite[Theorem 3.5]{LL15} is true. Therefore, it is
enough to prove that any LCD code over a finite commutative local
Frobenius ring $R_j$ is free.

\begin{Definition}
An $R_j$-module $C_j$ of rank $k$ is projective if there is an
$R_j$-module $M$ such that $(R_j)^k$ and $C_j\oplus M$ are
isomorphic (as $R_j$-modules).
\end{Definition}

\begin{Remark}\label{rem1}
Let $A_j$ and $B_j$ be $R_j$-modules. If $A_j\oplus B_j$ is free,
then $A_j$ and $B_j$ are projective.
\end{Remark}

\begin{Lemma}\cite[Theorem 2.]{Kap58}\label{lem2}
Any projective module over a local ring is free.
\end{Lemma}

In the following result, we prove that there does not exist
non-free LCD code over finite commutative local  Frobenius rings.

\begin{Theorem}\label{thm1}
Over finite commutative local Frobenius rings, any LCD code is
free.
\end{Theorem}

\begin{Proof}
Let $C_j$ be an LCD code over a commutative local Frobenius ring
$R_j$ and $n$ is the length of $C_j.$ Then $C_j\oplus C_j^\perp$
is a direct summand in $(R_j)^n.$ Since $R$ is Frobenius, by the
results in \cite{Woo99}, $C$ satisfies
$|C_j|\times|C_j^\perp|=|R_j|^n.$ Thereby $C_j\oplus
C_j^\perp=(R_j)^n.$ So the $R$-module $C_j\oplus C_j^\perp$ is
free, and by Remark \ref{rem1}, it follows that $C_j$ is
projective. Now $C_j$ is a finitely generated projective
$R_j$-module and $R_j$ is a local ring and by Lemma \ref{lem2},
$C_j$ is free.
\end{Proof}

 It follows from Theorem \ref{thm1} and \cite[Corollary 3.6]{LL15}  that there does not exist non-free LCD codes over finite commutative
local  Frobenius rings. We now show that the converse of
Theorem \ref{thm1} does not hold in general. To show this we cite
the following example.

\begin{Example}
Let $C$ be a linear code over $\mathbb{Z}_4$ with generator matrix
\[
\mathrm{G}=
  \left[ {\begin{array}{ccccccc}
   1 & 0 & 0  & 0 & 2 & 0 & 0   \\
   0 & 1 & 0  & 0 & 0 & 1 & 1  \\
   0 & 0 & 1  & 0 & 0 & 1 & 1  \\
   0 & 0 & 0  & 1 & 1 & 0 & 0  \\
   \end{array} } \right].\] Clearly $C$ is free. But $C$ is not LCD, as $(0, 0, 0, 2, 2, 0,0)\in C\cap C^\perp.$
\end{Example}


\begin{Proposition}\label{prop:gn}
A linear code $C_j$ over $R_j$ with generator matrix
$\mathrm{G}_j$. If $\mathrm{G}_j\mathrm{G}_j^T$ is nonsingular,
then $C_j$ is free.
\end{Proposition}

\begin{Proof} Suppose that $C_j$ is not free. Then $\mathrm{G}_j$ is not full-row-rank. From Lemma \ref{lem:full}, it
follows that $\mathrm{G}_j$ is not right-invertible. Hence
$\mathrm{G}_j\mathrm{G}_j^T$ is singular.
\end{Proof}

\begin{Corollary}\label{cor:sin}
A linear code $C_j$ over $R_j$ with generator matrix
$\mathrm{G}_j$ is LCD, if and only if $\mathrm{G}_j\mathrm{G}_j^T$
is nonsingular.
\end{Corollary}

\begin{Proof} Suppose that $C_j$ is LCD with rank $k,$ and $\textbf{c}\in C_j.$
From Theorem \ref{thm1}, $C_j$ is free and $\textbf{c}$ can be
written as $\textbf{c} = \textbf{v}\mathrm{G}$ for some
$\textbf{v}$ in $(R_j)^k.$ If $\mathrm{G}_j\mathrm{G}_j^T$ is
singular, by Corollary\,\ref{cor:ax} there is a nonzero vector
$\textbf{u}$ in $(R_j)^k$ such that
$\textbf{u}\mathrm{G}_j\mathrm{G}_j^T = \textbf{0}.$ Now
$\textbf{u}\mathrm{G}$ is a nonzero vector in $C_j.$ So that
$$(\textbf{u}\mathrm{G})\textbf{c}^T = (\textbf{u}\mathrm{G})(\textbf{v}\mathrm{G})^T
= \textbf{u}\mathrm{G}_j\mathrm{G}_j^T\textbf{v}^T =
\textbf{0}\textbf{v}^T = \textbf{0}$$ and hence
$\textbf{u}\mathrm{G}$ is also a vector in $C_j^\perp.$ It follows
that $C_j\cap C_j^\perp\neq\{\textbf{0}\},$ i.e., that $C$ is not
LCD. Absurd, therefore $\mathrm{G}_j\mathrm{G}_j^T$ is
nonsingular.

Suppose that $\mathrm{G}_j\mathrm{G}_j^T$ is nonsingular.  Let
$\textbf{c}\in C_j\cap C_j^\perp,$ by Proposition \ref{prop:gn},
$C_j$ is free. On the one hand, $\textbf{c}\in C_j$ implies that
there is $\textbf{u}\in (R_j)^k$ such that $\textbf{c} =
\textbf{u}\mathrm{G}_j.$ It follows that
\begin{align}\label{a1}\textbf{c}\mathrm{G}_j^T(\mathrm{G}_j\mathrm{G}_j^T)^{-1}\mathrm{G}_j
=
\textbf{u}\mathrm{G}_j\mathrm{G}_j^T(\mathrm{G}_j\mathrm{G}_j^T)^{-1}\mathrm{G}_j
=\textbf{u}\mathrm{G}_j =\textbf{c},\end{align} and the other
hand, $\textbf{c}\in  C_j^\perp,$  it follows that
$\textbf{c}\mathrm{G}_j^T=\textbf{0}.$ So
\begin{align}\label{a2}\textbf{c}\mathrm{G}_j^T(\mathrm{G}_j\mathrm{G}_j^T)^{-1}\mathrm{G}_j = \textbf{0}(\mathrm{G}_j\mathrm{G}_j^T)^{-1}\mathrm{G}_j =\textbf{0}.\end{align}
From (\ref{a1}) and (\ref{a2}), it follows that
$\textbf{c}=\textbf{0}.$ Whence $C_j$ is LCD.
\end{Proof}

\begin{Example} The linear code $C$ of length 8 generated by $G=\left[\begin{array}{cccccccc}
1&0&0& 0& 0& 1& 2& 1\\
0& 1& 0 &0 & 1& 2& 3& 1\\
0& 0& 1 &0& 0& 0& 3& 2\\
0& 0& 0& 1& 2 &3& 1& 1\\
\end{array} \right]$ over $\mathbb{Z}_4$ is LCD code whose minimum Lee distance is 4 and has free rank 4 ($[8,4^4,4]$-code). The Gay image of $C$ is a non-linear binary code of length 16 and minim Hamming distance 4.
\end{Example}

 A linear $[n,k]$-code $C'$ over $R_j$ is a \emph{lift} of a linear $[n,k]$-code $C$ over
$S$ by ring epimorphism $f_j:S\rightarrow R_j,$ if $C'=f_j(C),$
where
$$f_j(C):=\left\{(f_j(c_1),f_j(c_2),\cdots,f_j(c_n))\;:\;(c_1,c_2,\cdots,c_n)\in
C\right\}.$$ We call $C'$ the \emph{projection} of $C$ by $f_j. $

\begin{Lemma}\label{lem:timf} Let $S$ and $R_j$ be finite commutative local Frobenius
rings with $S^\times$ and $(R_j)^\times$ the unit group of $S$ and
$R,$ respectively. Then $f_j(S^\times)=(R_j)^\times,$ for any ring
epimorphism $f_j:S\rightarrow R_j.$
\end{Lemma}

The following result is a generalization of \cite[Theorem
3.9]{LL15} to any finite commutative local Frobenius ring $S$ and
any ring epimorphism $f_j:S\rightarrow R_j.$

\begin{Theorem}\label{thm:proj1} Let $S$ and $R_j$ be finite commutative local Frobenius
rings. The projection of any LCD $[n,k]$-code over $S$ by ring
epimorphism $f_j:S\rightarrow R_j,$ is also an LCD $[n,k]$-code
over $R_j.$
\end{Theorem}

\begin{Proof} Let $C$ be an LCD $[n,k]$-code over $S$ with a generator
matrix $\mathrm{G}.$ From Theorem\,\ref{thm1}, $C$ is free.
Therefore the projection $f_j(C)$ of $C$ by $f_j$ is a free
$[n,k]$-code over $R_j$ with a generator matrix $f_j(\mathrm{G}).$
Now
$$f_j\left(\texttt{det}(\mathrm{G}\mathrm{G}^T)\right)=\texttt{det}\left(f_j(\mathrm{G})f_j(\mathrm{G}^T)\right).$$
 From Lemma\,\ref{lem:timf} and Theorem\,\ref{cor:sin}, it follows that ,
$\texttt{det}\left(f_j(\mathrm{G})f_j(\mathrm{G}^T)\right)$ is a
unit in $R_j.$ Whence $f_j(C)$ is a LCD $[n,k]$-code over $R_j.$
\end{Proof}

The result revisits and extends \cite[Theorem 3.10]{LL15} to any
finite commutative local Frobenius ring $S$ and any ring
epimorphism $f_j:S\rightarrow R_j.$

\begin{Theorem}\label{thm:proj2} Let $S$ and $R_j$ be finite commutative local Frobenius
rings. Any lift of an LCD $[n,k]$-code over $R_j$ by ring
epimorphism $f_j:S\rightarrow R_j,$ is also an LCD $[n,k]$-code
over $S.$
\end{Theorem}

\begin{Proof} Let $C'$ be an LCD $[n,k]$-code over $R_j$ with a generator
matrix $\mathrm{G}_j.$ Since $f_j:S\rightarrow R_j$ is a
ring-epimorphism, there is a full-row-rank matrix $\mathrm{G}$
over $S$ such that $\mathrm{G}_j=f_j(\mathrm{G}).$ Consider the
free $[n,k]$-code $C$ over $S$ with generator matrix $\mathrm{G}.$
Now
$$f_j\left(\texttt{det}(\mathrm{G}\mathrm{G}^T)\right)=\texttt{det}\left(f_j(\mathrm{G})f_j(\mathrm{G}^T)\right).$$
By Lemma\,\ref{lem:timf} and Theorem\,\ref{cor:sin}, it follows
that $\mathrm{G}\mathrm{G}^T$ is nonsingular. Consequently, $C$ is
LCD.
\end{Proof}

 The map $$\begin{array}{cccc}
                    \pi_m: & R_m & \rightarrow & \mathbb{F}_2 \\
                           &   \sum\limits_{A\subseteq\{1,2,\cdots,m\}}c_A\prod\limits_{i\in A}u_i & \mapsto & c_\emptyset
                          \end{array} $$
is a ring-epimorphism. From Theorems\, \ref{thm:proj1} and
\ref{thm:proj2}, a linear code $C$ over $R_m$ is LCD, if and only
if $\pi_m(C)$ is a binary LCD code. From \cite[Theorem 1]{CMTQ18},
if $(1,1,\cdots,1)\not\in \pi_m(C)^\perp$ then $C$ is LCD if and
only if there exists a basis $\textbf{c}_1, \textbf{c}_2,\cdots,
\textbf{c}_k$ of $C$ such that $[\textbf{c}_i,
\textbf{c}_j]=\delta_{i,j},$ for all $1\leq i,j\leq k.$

\begin{Example} Consider the linear $[n,k]$-code $C$ over $R_m$ with generator matrix
$$\left(
\begin{array}{cccccccc}
1 & 0 & 0 & \cdots  & 0 & \lambda_{1,1} & \cdots  & \lambda_{1,n-k} \\
0 & 1 & 0 & \cdots  & 0 & \lambda_{2,1} & \cdots  & \lambda_{2,n-k} \\
\vdots  & \ddots  & \ddots  & \ddots  & \vdots  & \vdots  &  & \vdots  \\
0 & \cdots  & 0 & 1 & 0 & \lambda_{k-1,1} & \cdots  & \lambda_{k-1,n-k} \\
0 & \cdots  & 0 & 0 & 1 & \lambda_{k,1} & \cdots  &
\lambda_{k,n-k}
\end{array}%
\right),$$ where $n-k$ is an even integer and
$\pi_m(\lambda_{i,j})=1,$ for all $1\leq i \leq k, 1\leq j \leq
n-k.$ From Theorem \ref{thm:proj2}, the code $C$ is LCD, since
$\pi_m(C)$ is a binary LCD code, by \cite[Theorem 1]{CMTQ18}.
\end{Example}

\begin{Theorem}\label{thm}
 A linear code  $C = \texttt{CRT}(C_1,C_2, \cdots , C_u )$ over $R$  is
LCD  if and only if,  the linear code $C_j$ over $R_j$ is LCD, for
all $1 \leq j\leq u.$
\end{Theorem}

\begin{Proof} The map $\Phi:  R  \rightarrow   R_1\times\cdots\times
R_u$ is a ring-isomorphism, and by Lemma\,\ref{lem4}, it follows
that $$C\cap C^\perp=\texttt{CRT}(C_1\cap C_1^\perp,C_2\cap
C_2^\perp, \cdots , C_u\cap C_u^\perp ).$$ Thus $C$  is LCD  over
$R$ if and only if, $C_j$ is LCD over $R_j$ for all $1 \leq j\leq u.$
\end{Proof}

\begin{Remark}\label{rem}
 From Lemma\,\ref{lem3} and Theorem\,\ref{thm}, it is easy to see that an LCD code
$C = \texttt{CRT}(C_1,C_2, \cdots , C_u )$ over $R$ is non-free,
if and only if there are $1\leq j_1<j_2\leq u$ such that
$\texttt{rank}_{R_{j_1}}(C_{j_1})\neq\texttt{rank}_{R_{j_2}}(C_{j_2}).$
\end{Remark}

\begin{Example}\label{ex} Let $C_1$ be an LCD code over $\mathbb{Z}_3$ with generator
matrix $\mathrm{G}_1:=\left(%
\begin{array}{ccccc}
  1 & 0 & 0 & 1 & 1 \\
  0 & 1 & 0 & 1 & 1 \\
  0 & 0 & 1 & 1 & 1
\end{array}\right),$ and $C_2$ be an LCD code over $\mathbb{Z}_5$ with generator
matrix $\mathrm{G}_2:=\left(%
\begin{array}{ccccc}
  1 & 0 & 1 & 1 & 1 \\
  0 & 1 & 0 & 4 & 2
\end{array}\right).$ From Remark\,\ref{rem}, the Chinese product of $C_1$ and $C_2$ is the
non-free LCD code $C$ over $\mathbb{Z}_{15}$ with generator matrix
$$\mathrm{G}:=\left(%
\begin{array}{ccccc}
  1 & 0 & 6 & 1 & 1 \\
  0 & 1 & 0 & 4 & 7 \\
  0 & 0 & 10 & 10 & 10
\end{array}%
\right),$$ since $\texttt{rank}_{\mathbb{Z}_{3}}(C_1)=3\neq
2=\texttt{rank}_{\mathbb{Z}_{5}}(C_2).$ But $C$ is a projective
module over $\mathbb{Z}_{15}.$
\end{Example}

We now are ready to answer the question: \textsl{"Do non-free LCD codes over finite
commutative Frobenius ring $R$ exist?"}. It is evident from Example \ref{ex} that  \textsl{"non-free LCD codes over finite commutative Frobenius rings exist and they are projective
modules over $R.$"}

\section{Constacyclic LCD codes over finite chain rings}\label{sec:4}

 Throughout  this section $R$ will denote a finite chain ring (and hence a Frobenius ring)  with residue field
 $\mathbb{F}_q,$ $\gamma$ a unit in $R,$  and $n$ a positive integer relatively
prime to $q.$ The projection $\pi:R\rightarrow\mathbb{F}_q$
extends naturally to a projection $R[X]\rightarrow\mathbb{F}_q[X]$
as follows: $\pi(f)=\sum_i\pi(f_i)X^i$ for $f=\sum_i f_iX^i;$ also
a projection $R^n\rightarrow(\mathbb{F}_q)^n $ as follows:
$\pi(\textbf{c})=(\pi(c_0),\pi(c_1),\cdots,\pi(c_{n-1}))$ for
$\textbf{c}=(c_0,c_1,\cdots,c_{n-1}).$ Thus for any nonempty
subset $C$ of $R^n,$ $\pi(C)=\{\pi(\textbf{c})\;:\;\textbf{c}\in
C\}.$

Recall that a linear code $C$ of length $n$ over $R$ is
$\gamma$-\emph{constacyclic} if $(\gamma c_{n-1}, c_0, c_1, \cdots, c_{n-2}) \in C,$ whenever $(c_0, c_1,\cdots, c_{n-1}) \in C.$
$C$ is called cyclic and negacyclic, respectively,   when $\gamma$ is $1$
and $-1$. A constacyclic code of length $n$ over $R$
is \emph{non-repeated root} if $n$ and $q$ are coprime. It is
known that the $\gamma$-constacyclic codes over $R$ are identified
to ideals of $R[X]/\langle X^n-\gamma\rangle$ via the $R$-module
isomorphism
\begin{align*}
\begin{array}{cccc}
  \Upsilon : & R^n & \rightarrow & R[X]/\langle X^n-\gamma\rangle \\
    & (c_0, c_1,\cdots, c_{n-1}) & \mapsto & c_0+ c_1x+\cdots+
    c_{n-1}x^{n-1},
\end{array}
\end{align*}
where $x:=X+\langle X^n-\gamma\rangle.$ In this section, we deal
with non-repeated root $\gamma$-constacyclic LCD codes of length $n$ over $R.$

 Let
$k\in\{0,1,2,\dots,n\}$ and $f:=f_0+f_1X +\cdots+f_{k}X^k$ be a
polynomial of degree $\texttt{deg}(f):=k$ over $R$ with $k<n,$ we
denote by $\mathrm{M}_k(f),$ the $(n-k)\times n$-matrix defined
by:
\begin{align}
\left(
\begin{array}{cccccccc}
f_{0} & f_{1} & \cdots  & f_{k} & 0 & 0 & \cdots  & 0 \\
0 & f_{0} & f_{1} & \cdots  & f_{k} & 0 & \cdots  & 0 \\
\vdots  & \ddots  & \ddots  & \ddots  &  & \ddots  & \ddots  & \vdots  \\
0 & \cdots  & 0 & f_{0} & f_{1} & \cdots  & f_{k} & 0 \\
0 & \cdots  & 0 & 0 & f_{0} & f_{1} & \cdots  & f_{k}
\end{array}
\right).
\end{align} Obviously, if
$f_0$ is a unit in  $R,$ then the rank of $\mathrm{M}_k(f)$ is
$n-k.$ Note that for any free $\gamma$-constacyclic code $C$ over
$R$ of rank $n-k,$ there is an only monic polynomial $g$ of degree
$k$ dividing $X^n-\gamma$ in $R[X]$ whose $\mathrm{M}_{k}(g)$ is a
generator matrix for $C.$ This polynomial $g$ is called the
\emph{generator polynomial} of $C$ and the free
$\gamma$-constacyclic code over $R$ with generator polynomial $g$
of length $n,$ is denoted $\mathcal{P}(R;n;g).$ Conventionally,
$\mathcal{P}(R;n;g)=\{\,\textbf{0}\,\},$ if $\texttt{deg}(g)\geq
n.$ Thus $X^n-\gamma$ is the generator polynomial of
$\{\,\textbf{0}\,\}.$

From now on, $g$ denotes a monic polynomial over $R$ with $g(0)$
is a unit in $R,$ and the nonzero element $\gamma$ in $\Gamma(R)$
is the remainder of the Euclidian division of $X^n$ by $g.$

From \cite[Theorem 5.2.]{Lop13}, the quotient ring $R[X]/\langle\,
X^n-\gamma\,\rangle$ is a principal ideal ring, if either $R$ is a
field, or $X^n-\pi(\gamma)$ is free-square. Recall that a
polynomial over a finite field is called \emph{square-free}, if it
has no multiple irreducible factors in its decomposition.  Of
course, $X^n-\pi(\gamma)$ is free-square since
$\texttt{gcd}(n,q)=1.$ From \cite[Theorem 2.7]{NS00}, if $g\in
R[X]$ is monic and $\pi(g)$ is square-free, then $g$ factors
uniquely into monic, coprime basic-irreducible. For any polynomial
$f$ in $\mathbb{F}_{q}[X]$ dividing $X^{n} - \gamma$ \cite[Theorem
XIII.4]{McD74} implies the existence and unicity of a polynomial
$g\in R[X]$ such that $\pi(g)=f$ and $g$ divides $X^{n} -\gamma,$
since $X^{n} - \gamma$ is square-free in $\mathbb{F}_{q}[X].$ The
polynomial $g$ will be called the \emph{Hensel lift} of $f.$

\begin{Lemma}\cite[Lemma 3.1 (3)]{BFM18}\label{lem3} Let $g_1$ and $g_2$ be monic polynomials over $R$ dividing $X^n-\gamma.$ Then
\begin{align}\mathcal{P}(R;n;g_1)\cap\mathcal{P}(R;n;g_2)=\mathcal{P}(R;n;\mu(g_1,g_2)),\end{align}
where $\mu(g_1,g_2)$ denotes the Hensel lift of
$\texttt{lcm}(\pi(g_1),\pi(g_2))$
\footnote{$\texttt{lcm}(\pi(g_1),\pi(g_2))$: the least common
multiple of $\pi(g_1),\pi(g_2).$}  to $R[X].$
\end{Lemma}

For a polynomial $f(X)$ of degree $r,$  $f^\ast(X)$ denotes its
\emph{reciprocal polynomial} and is given by $f^\ast(X)=X^rf(\frac{1}{X}).$ A polynomial
$f(X)$ is \emph{self-reciprocal}, if $f^\ast(X)=f(X).$ Consider
the permutation $\Phi:R^n\rightarrow R^n$ defined as follows:
$\Phi((a_0, a_1, \dots , a_{n-1})) = (a_{n-1}, a_{n-2}, \dots ,
a_0).$ Recall that a linear code $C$ of length $n$ over $R$ is
\emph{reversible} if $\Phi(C)= C.$ Obviously,
\begin{align}\label{renver}\Phi(\mathcal{P}(R;n;g))=\mathcal{P}(R;n;g^\ast).\end{align} On the
other hand, for any $\gamma$-constacyclic code
$C=\mathcal{P}(R;n;g),$ it is well-know that
\begin{align} C^\perp=\Phi(\mathcal{P}(R;n;\widehat{g})), \end{align} where
 $\widehat{g\,}(X)g(X)=X^n-\gamma.$ This leads to the
following result.

\begin{Proposition}\label{dual}
Let $g$ be a monic polynomial $g$ of degree $k$ dividing
$X^n-\gamma.$ If $C=\mathcal{P}(R;n;g),$ then
$C^\perp=\mathcal{P}(R;n;\widehat{g^\ast\,}),$ where
$\widehat{g\,}(X)g(X)=X^n-\gamma.$
\end{Proposition}

From the precedent result, we have
$C^\perp=\mathcal{P}(R;n;\widehat{g^\ast\,})$ and
$\widehat{g^\ast\,}(X)$ divides $X^n-\gamma^{-1}.$ For this, we
have the following result.

\begin{Proposition}\label{dual:consta} The dual of any $\gamma$-constacyclic code
over $R$ is $\gamma^{-1}$-constacyclic.
\end{Proposition}

 Obviously, both $\{\,\textbf{0}\,\}$ and $R^n$ are $\gamma$-constacyclic codes for any unit
 $\gamma$ in $R.$ Inversely, we have the following result.

\begin{Lemma}\label{nega}
Let $C$ be a free code of length $n$ over $R.$ If $C$ is both
$\alpha$-constacyclic and $\beta$-constacyclic for $\alpha, \beta
$ units in $R$ with $\pi(\alpha)\neq\pi(\beta),$ then either $C =
\{\,\textbf{0}\,\}$ or $C = R^n.$
\end{Lemma}

\begin{Proof} Assume that $C\neq\{\,\textbf{0}\,\}.$ There exists a polynomial
$g:=g_0+g_1X+\cdots+g_{k-1}X^{k-1}+X^k$ with $k<n$ such that
$C=\mathcal{P}(R;n;g).$ Then the word
$\textbf{c}:=(0,\cdots,0,g_0,g_1,\cdots,g_{k-1}, 1)$ belongs to
$C.$ Since $C$ is both $\alpha$-constacyclic and
$\beta$-constacyclic, it follows that
$\textbf{c}_\alpha:=(\alpha,0,\cdots,0,g_0,g_1,\cdots,g_{k-1})\in
C$ and
$\textbf{c}_\beta:=(\beta,0,\cdots,0,g_0,g_1,\cdots,g_{k-1})\in
C.$ Thus
$\alpha\textbf{c}_\beta-\beta\textbf{c}_\alpha=(\alpha-\beta)(0,0,\cdots,0,g_0,g_1,\cdots,g_{k-1}).$
Now $\pi(\alpha)\neq\pi(\beta)$ and $C$ is linear over $R,$
therefore $\textbf{c}':=(0,0,\cdots,0,g_0,g_1,\cdots,g_{k-1})\in
C.$ And so on, we have $(0,\cdots,0,1)\in C$ since $g_0$ is a unit
in $R.$ By constacyclicity of $C,$ it follows that  $C=R^n.$
\end{Proof}

\begin{Corollary}
If $\pi(\gamma^2)\neq 1,$ then any free $\gamma$-constacyclic code
of length $n$ over $R$ is LCD.
\end{Corollary}

\begin{Proof}
Assume that $\pi(\gamma^2)\neq 1,$ and let $C$ be a free $\gamma$-constacyclic code of length $n$ over
$R.$  Then by Proposition \ref{dual:consta}, 
$C^\perp$ is a $\gamma^{-1}$-constacyclic code. Thus, $C \cap
C^\perp$ is both $\gamma$-constacyclic and
$\gamma^{-1}$-constacyclic. Therefore, by Lemma \ref{nega}, $C\cap
C^\perp = \{\,\textbf{0}\,\},$ i.e., $C$ is an LCD code as  because
$C\cap C^\perp$ can not be $R^n,$ when $\pi(\gamma^2)\neq 1$.
\end{Proof}

Thus, in order to obtain all $\gamma$-constacyclic  LCD codes, we
need to consider only the case when $\pi(\gamma^2) = 1.$ Moreover,
 the dual code of any $\pi(\gamma)$-constacyclic code over $\mathbb{F}_q$ is
still a $\pi(\gamma)$-constacyclic code over $\mathbb{F}_q$ when $\pi(\gamma^2)=1$.

\begin{Lemma}\label{chara1}
Let $C$ be an $\alpha$-constacyclic code of length $n$ over
$\mathbb{F}_q$ with $\alpha^2=1.$ The following assertions are
equivalent.
\begin{enumerate}
    \item $C$ is LCD;
    \item $g$ is self-reciprocal;
    \item $C$ is reversible.
\end{enumerate}
\end{Lemma}

\begin{Proof} Let $C=\mathcal{P}(\mathbb{F}_q;n;f).$
From Proposition\,\ref{dual},
$C^\perp=\mathcal{P}(\mathbb{F}_q;n;\widehat{f^\ast\,})$ and since
$\alpha^2=1,$ it follows from  Proposition\,\ref{lem3} that $$C\cap
C^\perp=\mathcal{P}(\mathbb{F}_q;n;\texttt{lcm}(f\,,\,\widehat{f^\ast\,})).$$
So $C$ is LCD if and only if
$\texttt{lcm}(f\,,\,\widehat{f^\ast\,})=X^n-\alpha.$ Since
$\texttt{deg}(\widehat{f})=n-\texttt{deg}(f)$ and
$\texttt{deg}(f)=\texttt{deg}(f)^\ast$, it follows that $f$ and
$\widehat{f^\ast\,}$ are coprime. Hence
$\texttt{lcm}(f\,,\,\widehat{f^\ast\,})=f\widehat{f^\ast\,}.$ As
$f\widehat{f\,}=X^n-\alpha$ and $\alpha^2=1,$ it follows that
$$f^\ast\widehat{f^\ast\,}=X^n-\alpha^{-1}=X^n-\alpha=f\widehat{f^\ast\,},$$ which is
equivalent to saying $f=f^\ast.$ From Eq. (\ref{renver}), $C$ is reversible if and only if $f=f^\ast$.
\end{Proof}

\begin{Remark}\label{rem*} Let $g$ be a monic polynomial in $R[X].$ Since $\mathcal{P}(R;n;g)=\biggl\{\textbf{c}\in R^n\;:\;g \text{ divides
} \Psi(\,\textbf{c}\,)\biggr\},$ it follows that
$\pi(\mathcal{P}(R;n;g))=\mathcal{P}(\mathbb{F}_{q};n;\pi(g)).$
\end{Remark}
From Remark\,\ref{rem*}, Theorems \ref{thm:proj1} and
\ref{thm:proj2}, we have

\begin{Lemma}\label{chara2}
Let $C$ be a $\gamma$-constacyclic code of length $n$ over $R.$
Then $C$ is LCD if and only if $\pi(C)$ is both
$\pi(\gamma)$-constacyclic and LCD .
\end{Lemma}

\begin{Theorem}\label{thm6}
Let $C$ be a $\gamma$-constacyclic code of length $n$ over $R$ and
$g$ its generator polynomial. Then $C$ is LCD and $\gamma^2=1$  if
and only if $C$ is reversible.
\end{Theorem}

\begin{Proof} Let $C=\mathcal{P}(R;n;g).$ From
Proposition\,\ref{dual},
$C^\perp=\mathcal{P}(R;n;\widehat{g^\ast\,}).$ Since $\gamma^2=1,$
it follows that $g^\ast$ divides $X^n-\gamma.$ It can use Lemma
\ref{lem3} and we have $C\cap
C^\perp=\mathcal{P}(R;n;\mu(g\,,\,\widehat{g^\ast\,})).$ Then $C$
is LCD and $\gamma^2=1$ if and only if
$\mu(g\,,\,\widehat{g^\ast\,})=X^n-\gamma,$ this implies that
$\mu(g\,,\,\widehat{g^\ast\,})=g\widehat{g^\ast\,}.$ Since
$g\widehat{g\,}=X^n-\gamma,$ it follows that $g^\ast=g.$ By
Equality\,(\ref{renver}), $C$ is reversible.

Conversely, if $C$ is reversible, then $\pi(C)$ is also reversible.
From Lemmas\,\ref{chara1} and \ref{chara2}, $C$ is LCD. Moreover
if $C$ is reversible, by Equality\,(\ref{renver}), we have
$g^\ast=g.$ But $g\widehat{g}=X^n-\gamma$ and
$g^\ast\widehat{g^\ast\,}=X^n-\gamma^{-1}.$ So
$X^n-\gamma=X^n-\gamma^{-1},$ because $g^\ast=g.$ Whence
$\gamma^2=1.$
\end{Proof}

 We now will provide some examples to demonstrate our results. We used the Magma Computer Algebra System \cite{Magma} in our computations. We have got some good codes, some optimal known codes and some new optimal codes over $\mathbb{Z}_4$ \cite{table}. 
\begin{Example} The factorization of $X^{7}-1$ over $\mathbb{Z}_4$ into a product
of basic irreducible polynomials over $\mathbb{Z}_4$ is given by
$$X^{7}-1=(X-1)(X^{3}+2X^{2}+X+7)(X^{3}+3X^{2}+2X+7).$$ Let
$f(X)=X^{3}+2X^{2}+X+7$ and $g(X)=X^{3}+3X^{2}+2X+7.$ From
Theorem\,\ref{thm6}, we have
\begin{itemize}
\item The cyclic code $\mathcal{P}(\mathbb{Z}_{4};7; (X-1))$ is LCD and reversible. This is  $[7, 4^6, 2]$ optimal code.
\item The cyclic code $\mathcal{P}(\mathbb{Z}_{4};7;f(X))$ is not LCD,
since $f(X)$ is not self-reciprocal.
\item The cyclic code $\mathcal{P}(\mathbb{Z}_{4};7;f(X)g(X))$ is LCD,
since $f(X)g(X)$ is self-reciprocal. This code has minimum Lee distance 7 but has only 4 codewords.
\end{itemize}
\end{Example}

Note that if $C$ is $\gamma$-constacyclic of odd length over
$\mathbb{Z}_{4},$ then $C$ is LCD if and only if $C$ is
reversible.

\begin{Example} The factorization of $X^{15}-1$ over $\mathbb{Z}_4$ into a product of basic irreducible polynomials over $\mathbb{Z}_4$ is given by
$$X^{15}-1=(X-1)(X^{2}+X+1)(X^{4}+X^{3}+X^{2}+X+1)(X^{4}+2X^{2}+3X+1)(X^{4}+3X^{2}+2X+1).$$ The self-reciprocal polynomials  and the LCD codes generated by those seld-reciprocal polynomials  are shown in the following  table:

$$
\begin{tabular}{|c|c|c|}
\hline
Generators (self-reciprocal) of LCD code $C$  &  $[n, 4^{k_1},d_L]$ & Remarks\\[1mm]
\hline
$g_1=X-1$ & $[15, 4^{14},2]$ & \\
\hline
$g_2=X^2+X+1$ & $[15, 4^{13},2]$ &\\
\hline

$g_3=(X-1)(X^2+X+1) $ & $[15, 4^{12},2]$ &\\
\hline

$g_4=X^{4}+X^{3}+X^{2}+X+1$ & $[15, 4^{11},2]$ &\\
\hline

$g_5=(X-1)(X^{4}+X^{3}+X^{2}+X+1)$  & $[15, 4^{10},2]$ &\\
\hline

$ g_6=(X^2+X+1)(X^{4}+X^{3}+X^{2}+X+1)$  & $[15, 4^{9},4]$ & Good \\
\hline

$ g_7=(X-1)(X^2+X+1)(X^{4}+X^{3}+X^{2}+X+1)$ & $[15, 4^{8},4]$ & \\
\hline

 $g_8=(X^{4}+2X^{2}+3X+1)(X^4+3X^2+2X+1) $ & $[15, 4^{7},3]$ &\\
\hline

$ g_9=(X-1)(X^{4}+2X^{2}+3X+1)(X^4+3X^2+2X+1)$  & $[15, 4^{6},6]$ & Good \\
\hline

 $g_{10}=(X^2+X+1)(X^{4}+2X^{2}+3X+1)(X^4+3X^2+2X+1) $ & $[15, 4^{5},3]$ &\\
\hline

 $g_{11}=(X-1)(X^2+X+1)(X^{4}+2X^{2}+3X+1)(X^4+3X^2+2X+1) $ & $[15, 4^{4},6]$ & \\
\hline
$ g_{12}=(X^{4}+2X^{2}+3X+1)(X^4+3X^2+2X+1)(X^4+X^3+X^2+X+1) $ & $ [15, 4^{3},5] $ & \\
\hline
$g_{13}=(x-1)(X^{4}+2X^{2}+3X+1)(X^4+3X^2+2X+1)(X^4+X^3+X^2+X+1) $ & $ [15, 4^{2},10]$  & Good \\
\hline
$g_{14}=(X^2+X+1)(X^{4}+2X^{2}+3X+1)(X^4+3X^2+2X+1)(X^4+X^3+X^2+X+1)$ & $[15, 4^{1},15] $ & \\
\hline
\end{tabular}$$
\end{Example}

\begin{Example} 
The factorization of $X^9-1$, $X^{17}-1$, $X^{31}-1$ and $X^{63}-1$ over $\mathbb{Z}_4$ into a product of basic irreducible polynomials are given by $$X^9-1=(X-1)(X^2+X+1)(X^6+X^3+1),$$
$$X^{17}-1=(X-1)( X^8 + 2X^6 + 3X^5 + X^4 + 3X^3 + 2X^2 + 1)(X^8 + X^7 + 3X^6 + 3X^4 + 3X^2 + X + 1),$$
  $$X^{31}-1=h_1h_2h_3h_4h_5h_6h_7,$$ where $h_1=(X-1)$, $h_2=(X^5 + 3X^2 + 2X + 3)$, $h_3=(X^5 + 2X^4 + X^3 + 3)$, $h_4=(X^5 + 2X^4 + 3X^3 + X^2 + 3X + 3)$, $h_5=(X^5 + 3X^4 + X^2 + 3X + 3)$,
   $h_6= X^5 + X^4 + 3X^3 + X + 3)$, $h_7=(X^5 + X^4 + 3X^3 + X^2 + 2X + 3)$, and $$X^{63}-1= g_1g_2 \dots g_{13},$$ where 
$g_1=(X - 1) $, $g_2=(X^2 + X + 1),$ $g_3=(X^3 + 2X^2 + X + 3),$ $g_4=(X^3 + 3X^2 + 2X + 3),$ $g_5=(X^6 + 2X^3 + 3X + 1),$ $g_6=(X^6 + X^3 + 1),$ $g_7= (X^6 + 2X^5 + 3X^4 + 3X^2 + X + 1),$ $g_8=(X^6 + 2X^5 + X^4 + X^3 + 3X + 1)$; $g_9=(X^6 + 3X^5 + 2X^3 + 1)$; $g_{10}=(X^6 + 3X^5 + 2X^4 + X^2 + X + 1),$ $g_{11}=(X^6 + 3X^5 + X^3 + X^2 + 2X + 1),$ $g_{12}=(X^6 + X^5 + X^4 + 2X^2 + 3X + 1),$ $g_{13}=(X^6 + X^5 + 3X^4 + 3X^2 + 2X + 1)$.  In the following table, we list cyclic LCD codes   over $\mathbb{Z}_4$ of different lengths and their generators. It is noted that some of the codes (which are LCD) are good known codes and some are new optimal codes over $\mathbb{Z}_4$ \cite{table}.\\

$$
\begin{tabular}{|c|c|c|}
\hline
Generators of $C$  &  $[n, 4^{k_1},d_L]$ & Remarks \\[1mm]
\hline
$(X-1)(X^6+X^3+1)$ & $[9, 4^{2}, 6]$ & Good \\
\hline
$(X^6+X^3+1)$ & $[9, 4^{3}, 3]$ & Good \\
\hline
$(X-1)(X^2+X+1)$ & $[9, 4^6, 2]$ & Good \\
\hline
$(X^8 + X^7 + 3X^6 + 3X^4 + 3X^2 + X + 1)$ &  $[17, 4^{9}, 7]$ & Optimal \\
\hline
$(X-1)(X^8 + X^7 + 3X^6 + 3X^4 + 3X^2 + X + 1)$ &  $[17, 4^{8}, 8]$ & Optimal \\
\hline

$h_1h_2h_3h_4h_4h_7$ & $[31, 4^{10}, 16]$ & Optimal \\
\hline
$h_2h_3h_5h_6$ & $[31, 4^{11}, 12]$ & Optimal \\
\hline
$h_1h_5h_6$ & $[31, 4^{20}, 8]$ & Optimal \\
\hline
$h_2h_3$ & $[31, 4^{21}, 6]$ & Optimal \\
\hline

$g_2g_3g_4g_5g_6g_7g_9g_{10}g_{12}g_{13}$ & $[63, 4^{13}, 36]$ & Optimal \\
\hline
$g_1g_3g_4g_5g_6g_7g_9g_{10}g_{12}g_{13}$ & $[63, 4^{14}, 34]$ & Optimal \\
\hline
$g_3g_4g_6g_7g_8g_{10}g_{11}g_{12}g_{13}$ & $[63, 4^{15}, 21]$ & Optimal \\
\hline
$g_1g_6g_7g_{10}g_{11}g_{12}g_{13}$ & $[63, 4^{20}, 18]$ & Optimal \\
\hline
$ g_1g_5g_6g_7g_9g_{13}$  & $[63, 4^{32}, 16 ]$ & Optimal \\
\hline
$g_1g_2g_7g_8g_{10}g_{11}g_{12}g_{13}$  & $[63, 4^{24}, 14]$ & Optimal\\
\hline

$ g_2g_3g_4g_6g_{10}g_{12}$ & $[63, 4^{37}, 12]$ & Optimal \\
\hline

$g_1g_2g_3g_4g_{10}g_{12}$  & $[63, 4^{42}, 10]$ & Optimal\\
\hline

$g_2g_5g_6g_7g_{9}g_{13}$  & $[63, 4^{31}, 9]$ & Optimal\\
\hline

$g_3g_4g_7g_8g_{11}g_{13}$  & $[63, 4^{33}, 7]$ & Optimal\\
\hline

$g_1g_8g_{11}$  & $[63, 4^{50}, 6]$ & Optimal\\
\hline
\end{tabular}$$
\end{Example}

\begin{Example} The factorization of $X^{15}-1$ over $\mathbb{Z}_8$ into a product
of basic irreducible polynomials over $\mathbb{Z}_8$ is given by
$$X^{15}-1=(X-1)(X^{2}+X+1)(X^{4}+X^{3}+X^{2}+X+1)(X^{4}+4X^{3}+6X^{2}+3X+1)(X^{4}+3X^{3}+6X^{2}+4X+1).$$ Out of $1$ and $X^{15}-1,$
there are $14$ self-reciprocal polynomials dividing $X^{15}-1$ in
$\mathbb{Z}_{8}[X]$ and they are:
$$
\begin{array}{c}
g_1=X-1 ;  \\
g_2=X^2+X+1 ; \\
g_3=(X-1)(X^2+X+1) ;  \\
g_4=X^{4}+3X^{3}+6X^{2}+4X+1 ;  \\
g_5=(X-1)(X^{4}+3X^{3}+6X^{2}+4X+1)  ;  \\
 g_6=(X^2+X+1)(X^{4}+3X^{3}+6X^{2}+4X+1) ;   \\
 g_7=(X-1)(X^2+X+1)(X^{4}+3X^{3}+6X^{2}+4X+1) ;  \\
 g_8=(X^{4}+4X^{3}+6X^{2}+3X+1)(X^4+3X^3+2X^2+1) ;  \\
 g_9=(X-1)(X^{4}+4X^{3}+6X^{2}+3X+1)(X^4+3X^3+2X^2+1) ; \\
 g_{10}=(X^2+X+1)(X^{4}+4X^{3}+6X^{2}+3X+1)(X^4+3X^3+2X^2+1) ; \\
 g_{11}=(X-1)(X^2+X+1)(X^{4}+4X^{3}+6X^{2}+3X+1)(X^4+3X^3+2X^2+1) ; \\
 g_{12}=(X^{4}+3X^{3}+6X^{2}+4X+1)(X^{4}+4X^{3}+6X^{2}+3X+1)(X^4+3X^3+2X^2+1) ; \\
 g_{13}=(X-1)(X^{4}+3X^{3}+6X^{2}+4X+1)(X^{4}+4X^{3}+6X^{2}+3X+1)(X^4+3X^3+2X^2+1) ; \\
 g_{14}=(X^2+X+1)(X^{4}+3X^{3}+6X^{2}+4X+1)(X^{4}+4X^{3}+6X^{2}+3X+1)(X^4+3X^3+2X^2+1) . 
\end{array}
$$
From Theorem\,\ref{thm6}, the nontrivial cyclic code
$\mathcal{P}(\mathbb{Z}_{8};15;g_i)$ over $\mathbb{Z}_{8},$ is
LCD,  for all $1\leq i\leq 14.$ Moreover
$\mathcal{P}(\mathbb{Z}_{8};15;g_i(\gamma X))$ is a nontrivial
$\gamma$-constacyclic LCD code over $\mathbb{Z}_{8},$ where
$\gamma\in\{3;5;7\},$ for all $1\leq i\leq 14.$ Hence there are 56
nontrivial constacyclic LCD codes of length $15$ over
$\mathbb{Z}_{8}$.
\end{Example}

\begin{Example} The factorization of $X^{9}-1$ over $\mathbb{Z}_8$ into a product
of basic irreducible polynomials over $\mathbb{Z}_8$ is given by
$$X^{9}-1=(X-1)(X^{2}+X+1)(X^6+X^{3}+1).$$
All three factors of $X^9-1$ over $\mathbb{Z}_8$ are self-reciprocal polynomials  in
$\mathbb{Z}_{8}[X]$ and hence all cyclic codes of length 9 over $\mathbb{Z}_8$ are LCD and so reversible.
$$
\begin{tabular}{|c|c|}
\hline
Generators of $C$  &  $[n, 4^{k_1},d_H]$  \\[1mm]
\hline
$(X-1)$ & $[9, 8^{8}, 2]$ \\
\hline
$(X^6+X^3+1)$ & $[9, 8^{3}, 3]$ \\
\hline
$(X^2+X+1)$ & $[9, 4^7, 2]$  \\
\hline
$(X-1)(X^6+X^3+1)$ &  $[9, 8^{2}, 6]$ \\
\hline
$(X-1)(X^2 + X + 1)$ &  $[9, 8^{6}, 3]$  \\
\hline
$(X^2+X+1)(X^6+X^3+1) $ & $[9, 8^{1}, 9]$ \\
\hline
\end{tabular}
$$
\end{Example}

\begin{Example} The Cyclic code $C$ of length 5 generated by $g(X)=X^2 + (3w + 2)X + 1$ over $GR(4, 2)$, where $GR(4,2)$ is the Galois  Extension of $\mathbb{Z}_4$ order 2 and $w$ is a root of the basic primitive polynomial $X^2+X+1$, is LCD code and its minimum Hamming distance is 3 ($[5, 16^3, 3]$).
\end{Example}

\section{Conclusion}

In paper, we have done an extensive study of LCD codes over finite
commutative  Frobenius rings. We have first corrected a wrong
result given in \cite{LL15} which in deed led to the claim that
"there do not exist non-free LCD codes over finite commutative
local Frobenius rings". We also answered the question posed in the title that there exists non-free LCD codes over finite commutative Frobenius rings but not over finite commutative local Frobenius rings. We have also obtained a necessary and
sufficient condition for any linear code over a finite commutative
Frobenius ring to be LCD.  We also characterized non-repeated root 
constacyclic LCD codes and revercible over finite chain rings and we found some new optimal codes over $\mathbb{Z}_4$ which are infact cyclic LCD codes over $\mathbb{Z}_4$.  

\section*{Acknowledgements}

The first author of the paper would like to Ministry of Human
Resource and Development India for support financially to carry
out this work. Third author is  partially funded by Spanish-MINECO MTM2015-65764-C3-1-P research grant.

\bibliographystyle{elsarticle-num}

\end{document}